\documentclass[pdflatex,sn-mathphys-num]{sn-jnl}% Math and Physical Sciences Numbered Reference Style
\usepackage{graphicx}%
\usepackage{multirow}%
\usepackage{amsmath,amssymb,amsfonts}%
\usepackage{amsthm}%
\usepackage{mathrsfs}%
\usepackage[title]{appendix}%
\usepackage{xcolor}%
\usepackage{textcomp}%
\usepackage{manyfoot}%
\usepackage{booktabs}%
\usepackage{algorithm}%
\usepackage{algorithmicx}%
\usepackage{algpseudocode}%
\usepackage{listings}%
\usepackage[latin1,utf8]{inputenc}%
\usepackage[T1]{fontenc}%
\usepackage[section]{placeins}%
\usepackage{color}
\definecolor{gray}{rgb}{0.4, 0.4, 0.4}
\definecolor{dkred}{rgb}{0.8, 0.0, 0.0}
\definecolor{dkgreen}{rgb}{0.0, 0.5, 0.0}
\definecolor{dkblue}{rgb}{0.0, 0.0, 0.5}

\theoremstyle{thmstyleone}%
\theoremstyle{thmstyletwo}%

\theoremstyle{thmstylethree}%

\begin{document}

\title[Electrically-pumped near-infrared VCSEL epitaxially-grown on Si]{Electrically-pumped near-infrared VCSEL epitaxially-grown on Si}


\author*[1]{\fnm{Guilhem} \sur{Almuneau}}

\author[1]{\fnm{Alexandre} \sur{Arnoult}}

\author[1]{\fnm{Karim} \sur{Ben Saddik}}

\author[2]{\fnm{Clara} \sur{Cornille}}

\author[1]{\fnm{Pierre} \sur{Gadras}}

\author[3]{\fnm{Armel} \sur{Descamps-Mandine}}

\author[1]{\fnm{Richard} \sur{Monflier}}

\author[1]{\fnm{Benjamin} \sur{Reig}}

\author[2]{\fnm{Mickael} \sur{Martin}}

\author[2]{\fnm{J\'er\'emy} \sur{Moeyaert}}

\author[2]{\fnm{Thierry} \sur{Baron}}

\affil*[1]{
%\orgdiv{LAAS}, 
\orgname{LAAS-CNRS, Universit\'e de Toulouse, CNRS, INSA}, 
%\orgaddress{\street{7, avenue du Colonel Roche}, 
\city{Toulouse}, 
\postcode{F-31400}, 
\country{France}}

%\affil[]{$^\dagger$ Now with: ISOM, Universidad Politécnica de Madrid, Spain}

\affil[2]{\orgname{University Grenoble Alpes, CNRS, CEA/LETI-Minatec, Grenoble INP, LTM}, \city{Grenoble}, \postcode{F-38054}, \country{France}}

\affil[3]{\orgdiv{Centre de Micro-Caract\'erisation Raimond Castaing}, 
\orgname{Universit\'e de Toulouse, CNRS UAR3623}, 
%\orgaddress{\street{Street}, 
\city{Toulouse}, \postcode{F-31400}, 
%\state{State}, 
\country{France}}


\abstract{The monolithic integration of compact laser sources onto silicon remains a critical bottleneck for the scalable development of silicon photonics. In particular, the direct epitaxial growth of near-infrared vertical-cavity surface-emitting lasers (VCSELs) on silicon has long been hindered by the simultaneous requirements of low crystalline defect density in the active region and smooth interfaces in the distributed Bragg reflector (DBR) mirrors, both essential to achieving high cavity quality factors and low lasing thresholds.

Here, we report the direct epitaxial integration of near-infrared VCSELs on silicon substrates using a two-step growth strategy specifically designed to suppress defect propagation while preserving mirror interface abruptness. This approach enables effective confinement of threading dislocations away from the active region and ensures high-reflectivity DBRs with excellent structural uniformity. Structural and optical characterizations reveal an epitaxial stack of high crystalline quality, leading to laser performance comparable to VCSELs commonly grown on GaAs substrates.

We demonstrate, for the first time, directly grown VCSELs on silicon wafers exhibiting threshold currents of only a few milliamperes. These results establish a decisive proof of concept for the monolithic integration of good-performance VCSELs onto silicon platforms and open a viable pathway toward fully integrated, low-cost silicon-based photonic systems.}


\maketitle

\section{Introduction}\label{sec1}

The development of silicon-based laser sources represents a holy grail in the evolution of photonics, largely due to the difficulty of achieving stimulated emission in silicon, owing to its indirect bandgap. Nevertheless, silicon has become the leading platform for photonic integrated circuits (PICs) \cite{Shekhar2024} , driving significant research into the integration of III-V semiconductors such as GaN \cite{Semond2015}, GaSb \cite{Silvestre2025}, and InP \cite{Besancon2021} onto silicon for applications in microelectronics and optoelectronics.

Furthermore, Vertical-Cavity Surface-Emitting Lasers (VCSELs) have become particularly well-suited sources for photonic applications, ranging from datacom links to integrated sensing. They offer unique advantages for an expanding range of applications, meeting operational specifications with their very small footprint inherent to their vertical emission, high power conversion efficiencies, and low power consumption. This makes them suitable for 2D laser arrays, vertical integration and wafer-scale manufacturing processes, as well as mass production at relatively low cost. They also provide versatile, symmetric beam profiles and high modulation bandwidths for fast data transmission, thanks to their low threshold and strong temperature stability \cite{Qiu2025}.

Thanks to these strengths, a range of VCSEL-based applications has recently emerged, including photonic neural networks, vortex beam emitters, holographic devices, beam deflectors, atomic sensors, and biosensors.
To address the challenge of VCSEL integration on silicon substrates and photonic circuits, several approaches are being explored. The most pragmatic and industrially mature solution is hybrid integration, which assembles different materials on the same wafer or chip \cite{Haglund2015, Goyvaerts2021}. Flip-chip bonding and micro-transfer printing techniques allow dense integration of PICs with sufficient alignment tolerance of III-V dies for coupling to SiN waveguides. However, these processes are complex and costly, and the stringent requirements limit throughput when applied to large surfaces. Despite these constraints, heterogeneous integration has already been adopted by certain photonic foundries.

The historical approach of direct epitaxial growth of III-V materials on IV-based substrates was initiated in the 1980s\cite{Kroemer1987}. Most research has focused on direct growth of III-V lasers in a planar waveguide geometry. This has led to highly mature approaches and laser performance approaching that of lasers grown on lattice-matched substrates \cite{Liang2010}, as well as demonstrations of monolithic integration on silicon photonic circuits \cite{Dong2024}.

Notable recent demonstrations of direct growth on much more complex multilayer components have recently been reported, such as for quantum cascade lasers \cite{Nguyen-Van2018}. These structures, which require ultimate quality at the \r{a}rmströng scale in terms of interfaces and layer roughness, have been achieved by direct epitaxy on a silicon platform, without noticeable degradation in performance due to the heteroepitaxy of III-V materials on a non-polar substrate.

Pioneering works on the direct growth of VCSEL on Si substrate were carried out by two groups in the 1990s \cite{Deppe1990, Egawa1995}, but resulted in a large number of lattice defects, such as dislocations, anti-phase domains (APDs), and stacking faults, leading to poor optical properties of the distributed Bragg reflectors (DBR), and thus inhibiting lasing operation. More recent results have been reported on VCSEL growth on Ge substrates, which offer the advantages of large surface wafer availability, a lower price than III-V wafers, and which could be compatible with integration onto a Si platform \cite{Chen2024}. Since the AlGaAs alloy system is nearly lattice-matched with the Ge, good performances could be achieved on VCSEL devices \cite{Wan2024} despite the drawbacks of the polar-on-nonpolar epitaxy \cite{Kroemer1987}. Meanwhile, the monolithic epitaxy of high-quality III-V DBRs on Si remains a major hurdle, despite the attempts to reduce the threading dislocation density \cite{Tsuji1999,Guo2023,Laryn2025}.       

In this paper we report electrically-pumped laser operation at room temperature of a monolithically epitaxially grown VCSEL on a silicon (100) substrate. The good structural and crystalline quality through the entire AlGaAs-based multilayer stack, including the DBRs and multi-quantum well active region, is achieved by a two-step epitaxial sequence, first by metal organic chemical vapor deposition (MOCVD) for the nucleation and GaAs buffer layers, followed by molecular beam epitaxy (MBE) for the VCSEL vertical stack. The resulting optical and electrical properties of the VCSEL, grown heterogeneously on silicon, enable highly reflective DBRs and high gain from the active layer. This allows for electrically pumped continuous-wave lasing operation at room temperature with a threshold of only a few milliamperes.

\section{Results}\label{results}

\subsection{Hetero-epitaxy of III-V VCSEL structure on silicon}

A GaAs/Si template was first grown directly on a 300\,mm Si wafer by MOCVD, initially with the growth of an antiphase-boundary-free GaAs buffer layer on an on-axis Si(001) substrate, following the method developed by Martin et al. \cite{Martin2016}.
This method includes the growth of InGaAs/GaAs multi quantum wells combined with thermal cycle annealing (see Methods).

The VCSEL structure was regrown by MBE (Fig. S1 in the supplementary document), on top of the MOCVD GaAs/Si template, after first cutting the 300-mm silicon wafer into 2-inch wafers that are better suited to the MBE chamber configuration. The MBE growth conditions are described in the Methods section. The 940\,nm-VCSEL epitaxial design includes a 35-periods N-doped bottom DBR, a one-$\lambda$ cavity including three In$_{0.1}$Ga$_{0.9}$As/GaAs multi-quantum-wells (MQW), and a 17 periods P-doped top DBR. Below the VCSEL structure, a 1\,$\mu$m-thick highly N-doped GaAs buffer layer has the dual role of smoothing the surface after regrowth on the MOCVD-grown virtual substrate and creating a contact layer for electrical injection. Each DBR consists of alternating layers of Al$_{0.9}$Ga$_{0.1}$As and GaAs with a high refractive index contrast and very low lattice mismatch, centered to achieve maximum reflectivity around 940 nm. In order to optimize electrical conductivity of the N- and P-doped DBRs by forming smooth transitions between the conduction and valence bands at hetero-interfaces, a thin intermediate 20-nm-thick layer with a gradual Aluminum composition is inserted at each interface of the DBR\cite{Peters1993}. This gradient is achieved in the form of superlattices of variable period (digital-alloy) and does not affect the optical properties of the DBR.
A separate confinement heterostructure (SCH) based-cavity is defined, surrounding the MQWs, to confine injected carriers and control the optical field with a graded index layer (GRIN-SCH), which offers reduced resistance to carrier flow for electrical and optical confinement.  
Above the cavity, a 30-nm Al$_{0.98}$Ga$_{0.02}$As layer was
included to form a laterally confining aperture by selective oxidation. Above the top DBR a highly P-doped GaAs 50-nm cap layer is finally grown to serve as the top contact electrode.

\subsection{Morphology assessment by in-situ and ex-situ structural characterization means}

\begin{figure}[htbp]
\centering
\includegraphics[width=1\textwidth]{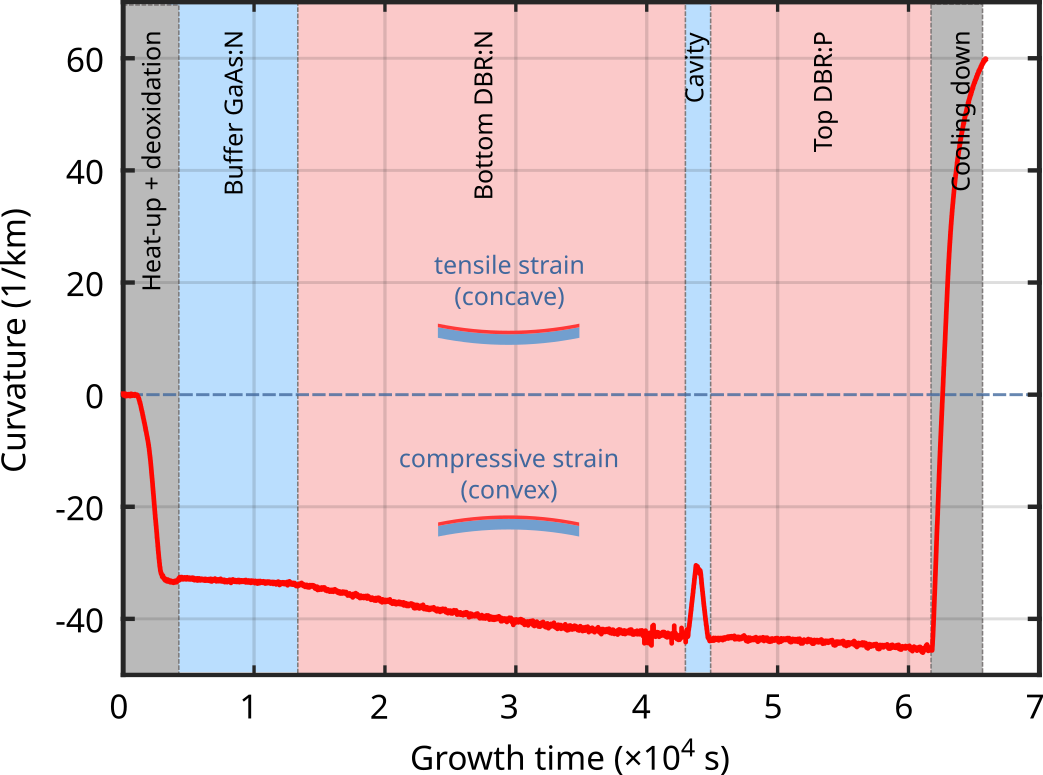}
\caption{In situ measurement of the curvature change during MBE growth.}
\label{fig:EZ_curve_strain}
\end{figure}

Throughout the MBE growth process, we monitored the evolution of the epitaxial layer parameters in real time using several in situ measurements, including wafer magnification-inferred curvature (Fig. \ref{fig:EZ_curve_strain}) and spectrally-resolved optical reflectometry (Fig. \ref{fig:EZ-ref_FTIR}a).
These measurements provide precise, real-time information on the growth process and any drifts and events that may occur in the morphology of the structure during epitaxy.
This information is essential in the case of highly mismatched and/or constrained growth, as here, as it allows us to identify relaxation or surface roughening effects at a given point in the growth process. Furthermore, these measurements provide quantitative information on the thicknesses, alloy compositions and characteristic wavelengths of the VCSEL component in real time.

Figure \ref{fig:EZ_curve_strain} shows the curvature evolution of the 2-inch Si wafer during the full MBE growth sequence of the VCSEL on a GaAs/Si template.

\FloatBarrier

The wafer curvature, measured in situ on an unclamped substrate, provides a quantitative assessment of film stress. This stress originates from thermal expansion mismatch between substrate and epilayers, as well as from epitaxial strain during growth~\cite{Arnoult2021}.

A significant curvature variation is observed during the initial heat-up and deoxidation stages, reaching about $-30\,\mathrm{km^{-1}}$ (concave), reflecting thermal stress in the GaAs/Si virtual substrate. A reverse trend occurs during cooling, with a larger amplitude after VCSEL growth, reaching $+60\,\mathrm{km^{-1}}$ at $\sim 93^\circ\mathrm{C}$. The difference between initial and final curvature corresponds to a wafer bow of $19.3\,\mu\mathrm{m}$ for the 2-inch wafer.

Overall, compressive strain accumulated during growth is largely compensated by tensile thermal strain upon cooling. This balance arises because the III--V layers are fully relaxed on Si, unlike VCSELs grown on GaAs, where these contributions do not cancel.

After the initial transient, curvature evolves linearly with time, reflecting stress-thickness behavior. During GaAs buffer growth, the nearly constant slope confirms a relaxed lattice. In contrast, both DBRs exhibit similar slopes to those grown on GaAs substrates, indicating that compressive strain arises from lattice mismatch between AlGaAs and GaAs without relaxation. The slightly reduced slope in the top DBR suggests lower strain, likely due to carbon-induced tensile contributions~\cite{Giannini1993}.

During cavity and MQW growth, abrupt curvature changes are observed, attributed to the reduced growth temperature required for optimal optical quality.

%%%%%%%%%%%%%%%%%%%%%%%%%%%%%%%%%%%%%%%%%%%%%%%%%%%%%%%%%%%%%

\begin{figure}[htbp]
\centering
\includegraphics[width=1\textwidth]{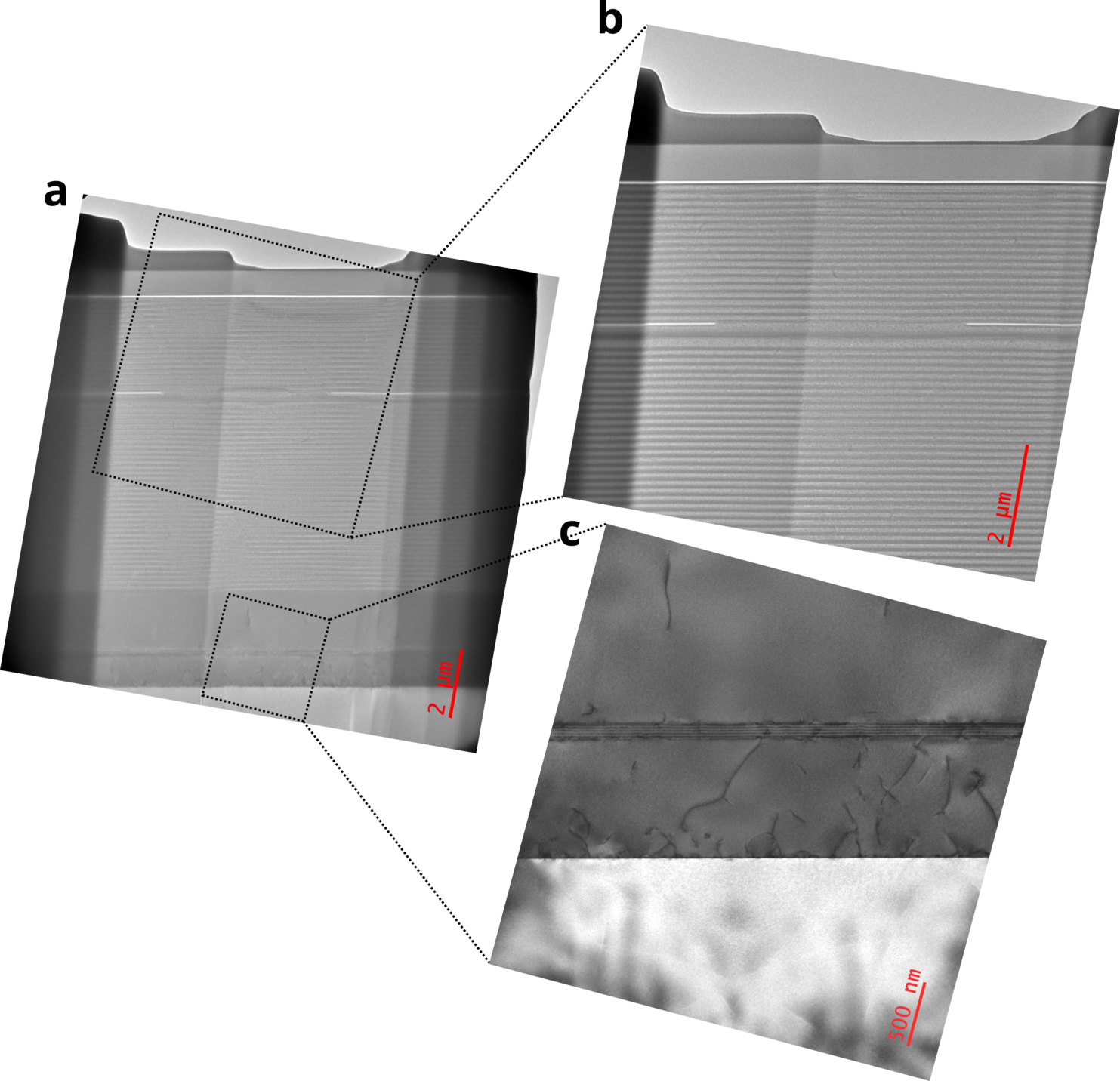}
\caption{ Cross-sectional TEM images showing: (a) the complete epitaxial stack, (b) the VCSEL structure grown by MBE, and (c) the buffer layer grown directly on silicon by MOCVD. The oxidized regions at each side of the aperture appear as white areas above the VCSEL cavity. The section above and including the white solid line at the top of the (a) and (b) images corresponds to materials introduced during focused ion beam (FIB) sample preparation.}\label{fig:TEM}
\end{figure}

The indicators of good structural properties observed during MBE growth are confirmed by transmission electron microscopy (TEM) of the cross-section of the full epitaxial stack.
As shown in Fig. \ref{fig:TEM}a-c, the VCSEL structure is free of threading dislocations within the observation zone, and the DBR multilayers and heterotructures appear abrupt and smooth. The vertical lines and changes in grey tone correspond to different thicknesses during the preparation of the lamellae by focused ion beam (FIB) etching.

\FloatBarrier

A close-up view of the bottom part of the epitaxy depicting the GaAs MOCVD growth directly on Si (Fig. \ref{fig:TEM}c) shows a high density of dislocations rapidly decreasing across the nucleation layer, and even more drastically vanishing above the superlattice.  
Furthermore, it can be noted that the regrowth interface between the part grown by MBE above the MOCVD template is not visible.
This cross-sectional view was taken after the end of VCSEL device fabrication process. Consequently, the oxide aperture created by selective thermal oxidation above the cavity is clearly visible, with the laterally oxidized layer from the edges of the mesa appearing as white lines (Fig. \ref{fig:TEM}b). 

\begin{figure}[htbp]
\centering
\includegraphics[width=1\textwidth]{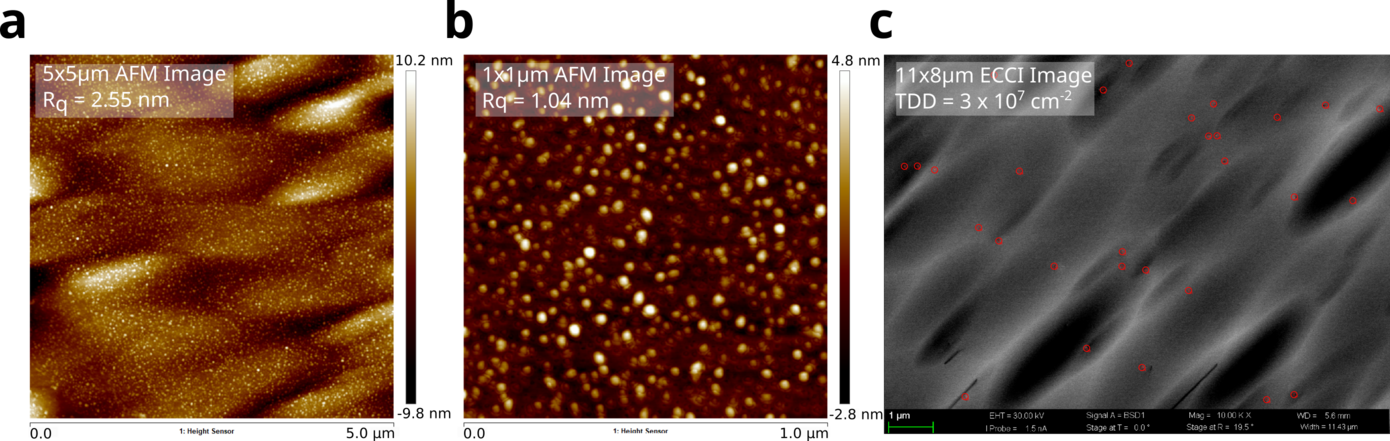}
\caption{(a-b) Two different scanning size AFM images, and (c) electron channelling image (ECCI) at the VCSEL surface. Threading dislocations are evidenced with small red circles on the ECCI image.}\label{fig:AFM-ECCI}
\end{figure}

A surface analysis by atomic force microscopy (AFM) was performed (see Figs. \ref{fig:AFM-ECCI} a and b) and indicates an average roughness of around 1 to 2.5\,nm depending on the analyzed area. Although this value is higher than in the case of lattice-matched epitaxy on a GaAs with roughness mean square (RMS) $\approx 0.2\,nm$, it remains similar to that achieved on Ge substrate (1-2\,nm) \cite{Wan2024, Haberland2003}. Nevertheless, these demonstrated values for VCSEL on Si remain fairly low considering the significant lattice mismatch between the Si substrate and the AlGaAs layers, especially in the case of a complete VCSEL structure including numerous interfaces and heterostructures.

\FloatBarrier

Electron channeling contrast imaging (ECCI) performed on the Si-VCSEL surface reveals a threading dislocation density (TDD) of $3\times10^{7}\,\mathrm{cm^{-2}}$ averaged over an analyzed area of $11\times8\,\mu\mathrm{m}^2$ (see Fig. \ref{fig:AFM-ECCI}c).

For comparison, previous studies have reported surface roughness and TDD values for DBR structures. Laryn \textit{et al.} reported an RMS roughness of 5.59~nm and a TDD of $1.28\times10^{8}\,\mathrm{cm^{-2}}$ for a 20-period InGaAs/AlGaAs DBR. Guo \textit{et al.} and Li \textit{et al.} employed GaAsP-based DBRs, achieving reduced TDDs in the $1.5$--$2.2\times10^{7}\,\mathrm{cm^{-2}}$ range, albeit with higher RMS roughness values of approximately 8.5--9.5~nm.

After epitaxy, optical microscopy inspection of the surface of the epitaxial VCSEL  reveals faint linear features aligned preferentially along a single crystallographic direction (the exact orientation could not be determined because the wafer orientation was not tracked). These lines are relatively difficult to observe, with a spacing ranging from roughly one hundred to several hundred micrometers. Their morphology and periodicity are consistent with thermally induced cracking, which is frequently reported in III-V epitaxial layers grown on silicon substrates due to the combined effects of thermal expansion mismatch and residual strain accumulation during growth and cooldown \cite{Yang2003}. Typically, for GaAs on Si, the critical thickness for the stress releasing and resulting thermal cracks formation is around 5-6$\,\mu$m for growth temperatures in the 550-600$^\circ$C range, which is well below than the 11.3$\,\mu$m total thickness of the III-V stack considered here.

During the subsequent device fabrication process, these linear defects became more apparent, particularly after the plasma etching steps. The etching process tends to widen pre-existing surface discontinuities, thereby increasing their optical contrast and making them easier to detect under microscopy. Even if these linear cracks were generally spaced out several hundreds of $\,\mu$m apart, their presence on the surface of the epitaxial wafer reduced the number of functional components. 

Cross-sectional transmission electron microscopy (TEM) observations confirmed that these lines correspond to cracks propagating orthogonally to the layers plane through the entire thickness of the VCSEL epitaxial stack and extending down toward the silicon substrate, where they gradually terminate (Fig. S4). The cracks appear nearly vertical with respect to the wafer surface and do not show significant lateral deviation along the multilayer interfaces. Importantly, the TEM analysis did not reveal noticeable degradation of the crystalline quality of the surrounding epitaxial layers, suggesting that the cracks primarily act as mechanical stress-relief features rather than sources of extended crystalline disorder within the stack.

We conducted high-resolution x-ray diffraction (HR-XRD) by measuring $\theta$-$2\theta$ diffractograms, as well as reciprocal space maps (RSM), on both the GaAs/Si templates and on full VCSEL-on-Si epitaxy samples. The objectives were to investigate the degree of relaxation of the III-V layers and to quantify the crystalline quality of the VCSEL epitaxial stack \cite{Alaydin2019}.

\begin{figure}[htbp]
\centering
\includegraphics[width=1\textwidth]{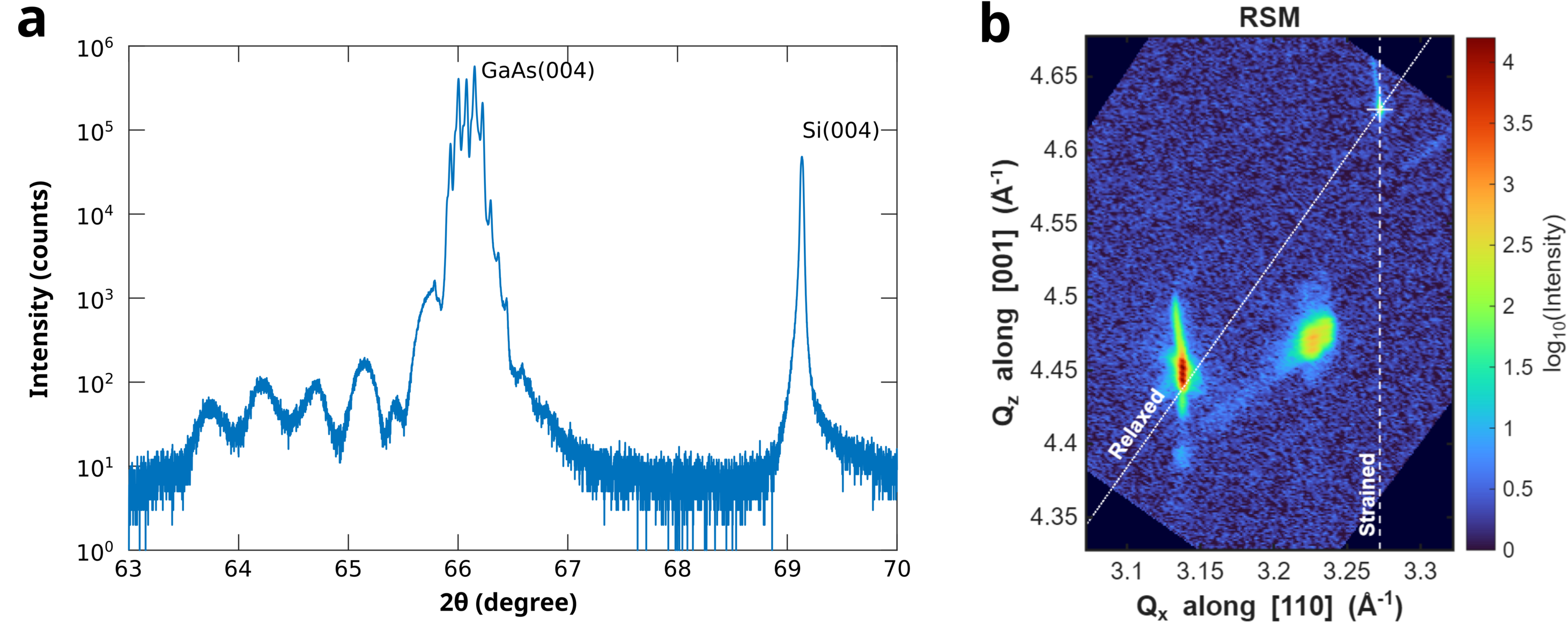}
\caption{(a) $\theta$-$2\theta$ measured diffractogram of (004) plane, and (b) Reciprocal space map around (224)+ reflection of the Si substrate (white cross). The map shows fully relaxed peaks related to fully relaxed VCSEL structure.}\label{fig:XRD}
\end{figure}

Figure \ref{fig:XRD}(a) presents the (004) $\theta$-$2\theta$ linescan measured on the full VCSEL structure. The peaks related to the AlGaAs-VCSEL structure are clearly separated from the Si substrate, indicating complete relaxation. In addition, a series of narrow peaks including the GaAs peak at 66$^\circ$ are visible with the high number of replicas inherent to the periodic nature of the DBR stacks. 
The presence of these sharp peaks and the ability to measure a large number of satellites unambiguously indicates that the interfaces within the DBRs are very even and have excellent periodicity.

At lower diffraction angles (below 66$^\circ$), several peaks with intensities less than one-thousandth of the later main peaks have also been measured in diffractograms taken from GaAs/Si template samples grown by MOCVD (Fig. S3). Indeed, they may be associated with the strongly dislocated layers in the nucleation region located just above the Si/GaAs interface or possibly due to the distorted growth zone near the thermal cracks. These partially relaxed layers are also observed in RSM measurements performed on the VCSEL structure presented just after, and on GaAs/Si template samples(Fig. S3).

Figure \ref{fig:XRD}(b) shows the RSM measured on the same VCSEL sample. Three different groups of peaks appear: the Si substrate; the fully relaxed III-V layers, as seen on the  $\theta$-$2\theta$ diffractogram; and a third group, which is located at an intermediate lattice parameter between the relaxed AlGaAs group and the Si substrate and is about 100-fold less intense described before.  Returning to the AlGaAs VCSEL-related peaks, we can draw a similar conclusion to that of the $\theta$-$2\theta$ scan: the fully relaxed lattice and multiple satellite peaks are present along a vertical line (same $Q_x$ value).

\FloatBarrier

Notably, the present results are obtained on a fully epitaxial VCSEL rather than a standalone DBR, yet they exhibit substantially lower surface roughness and a TDD comparable to the best state-of-the-art DBRs directly grown on Si \cite{Tsuji1999, Balakrishnan2006, Guo2023, Laryn2025}.

These findings further confirm the low dislocation densities achievable using GaAs/Si templates grown by MOCVD, consistent with earlier demonstrations on III-V resonant-cavity micro-LEDs integrated on large-area Si substrates~\cite{Hijazi2025}.

\subsection{Optical properties}

\begin{figure}[htbp]
\centering
\includegraphics[width=1\textwidth]{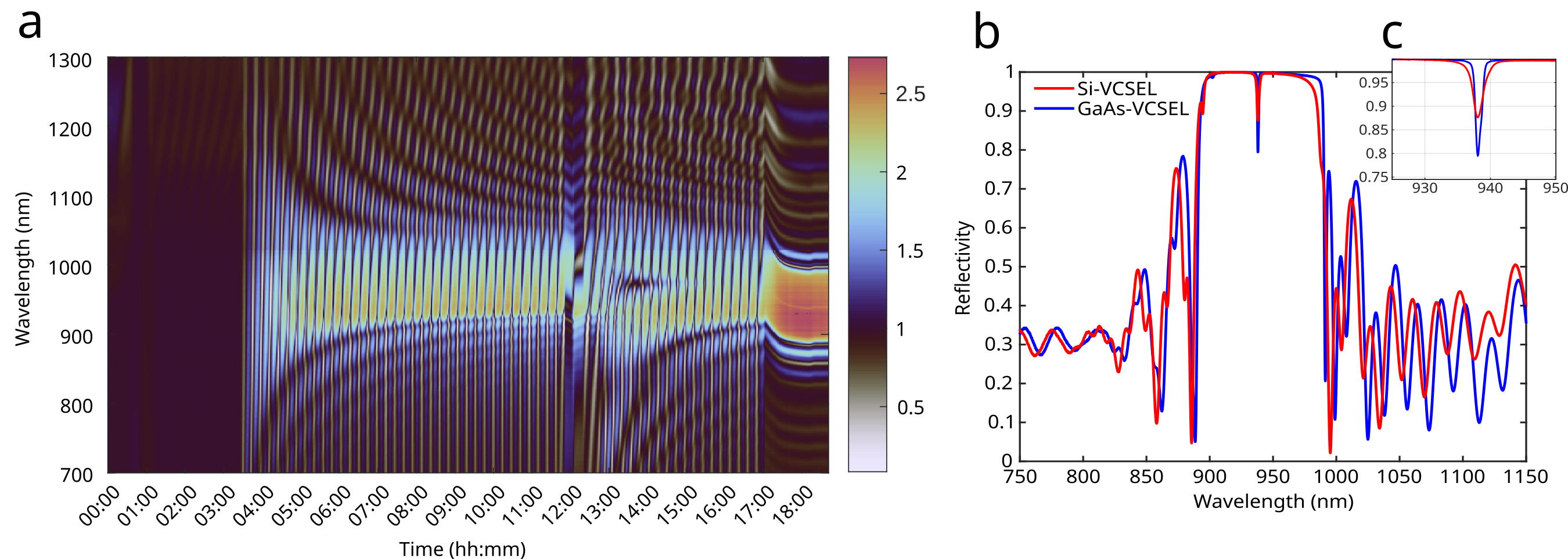}
\caption{(a) In-situ measurement of the reflectivity spectrum of the VCSEL structure throughout the MBE growth process. (b) ex-situ post-growth reflectivity measurement by FTIR at normal incidence of the VCSEL grown on Si (red) and on a VCSEL grown on GaAs (blue) taken as a reference. (c) Zoom-in on the cavity resonance peak. The spectrum of the GaAs-VCSEL was intentionally shifted in wavelength to match with the resonance of the Si-VCSEL.}\label{fig:EZ-ref_FTIR}
\end{figure}

In this section, we present the optical properties of the VCSEL microcavity structure.
First, we present the temporal evolution of the reflectivity spectrum during the MBE growth of the VCSEL.
Figure \ref{fig:EZ-ref_FTIR}a illustrates this evolution in the form of a colormap, showing the reflectivity spectrum at normal incidence as the VCSEL structure grows at the stabilized growth temperature around 600$^{\circ}$C.
The DBRs are clearly identifiable due to their multilayer nature. As the number of layers increases, interference fringes form, strengthening the high reflectivity stopband around the central wavelength and gradually producing secondary partial interference lobes for wavelengths below and above the stopband.  

\FloatBarrier

Following the growth of the cavity around the 12:00 timestamp, a dip in reflectivity at the microcavity resonance becomes clearly visible at around 975 nm during the growth of the upper DBR. This resonance narrows as the number of periods increases and the quality factor of the optical cavity improves. The resonance observed at 975 nm at the growth temperature of 600$^{\circ}$C allows us to predict a resonance dip at 935-940 nm at room temperature.
This shift in the dip wavelength is also observed during the temperature decrease phase after growth ends (after timestamp 17:00).

These in-situ observations during the growth are confirmed by ex-situ reflectivity measurements taken at normal incidence and at room temperature using Fourier transform infrared spectroscopy (FTIR) (see Figs. \ref{fig:EZ-ref_FTIR}b and c). The reflectivity curve of the VCSEL grown on Si shows a well-defined a broad plateau of very high reflectivity (stopband). The reflectivity spectrum of the Si-VCSEL is compared to that of a  reference VCSEL homo-epitaxially grown on GaAs and the responses are found to be very similar.

Absolute reflectivity cannot be accurately measured using a single point measurement with FTIR. However, the width of the stopband equivalent to a VCSEL on GaAs, and based on calculations from ref. \cite{Khreis2016}, enables us to conclude that the reflectivity of the DBRs exceeds 99.5\%. Figure \ref{fig:EZ-ref_FTIR}c shows a zoom of the resonance peak with a half-width of 1.8 nm, compared to 1.1 nm for GaAs.
These values, which are related to the cavity quality factor (Q-factor), should be interpreted with caution, since they depend also on the spectral resolution of the measurement and on the extent to which the spectral overlap coincides with the position of the quantum well gain curve. Cavity losses related to the half-height width are greater when the cavity wavelength overlaps with a high value on the gain curve.

These results on the optical properties of GaAs/AlGaAs DBRs on Si are clearly superior to those of recent studies \cite{Li2025, Guo2023, Laryn2025}. In the latter, the presence of non-abrupt interfaces (e.g. due to significant roughness) reduces the stop-band width and damps the sidelobe interference fringes on either side of the stopband (see \cite{Khreis2013}). 

\subsection{Wafer bow measurement by deflectometry} \label{Deflecto}

VCSELs require thick epitaxial stacks due to the large number of mirror pairs in the DBRs, leading to substantial built-in compressive strain. This induces wafer bowing, which complicates large-area fabrication through photolithography non-uniformity, oxidation issues, and increased risk of wafer breakage. Although backside silicon nitride deposition can partially compensate for bowing, it adds process complexity and time.

On large-diameter GaAs wafers (150--200\,mm), bowing exceeding 150\,$\mu$m is commonly observed despite the low lattice mismatch of AlGaAs alloys. Using Ge substrates mitigates this effect due to their intermediate lattice parameter, which enables partial strain compensation between DBR layers.

\begin{figure}[htbp]
\centering
\includegraphics[width=0.5\textwidth]{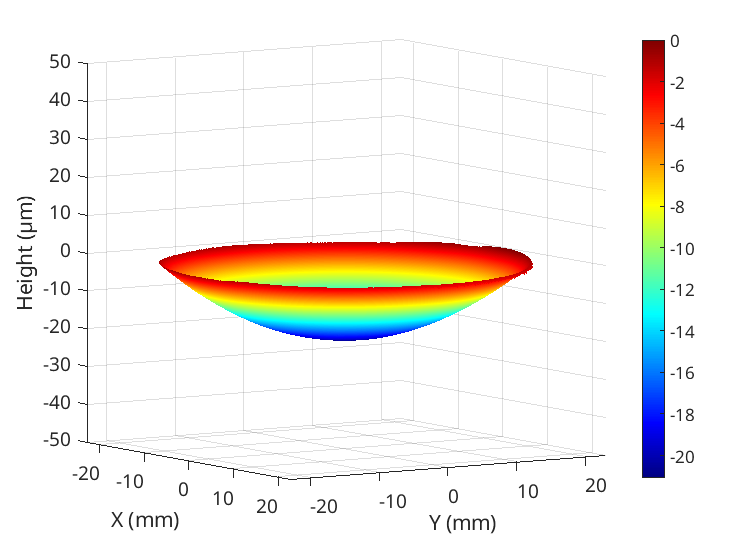}
\caption{Reconstructed height map from deflectometry measurements on a 2-inch Si-based VCSEL wafer.}
\label{fig:Deflecto}
\end{figure}

For Si-based VCSELs, strain-induced bowing remains largely unexplored but is critical for scaling to microelectronics-compatible wafer sizes. Post-growth deflectometry measurements reveal a concave bow of 21.4\,$\mu$m on a 2-inch Si substrate (thickness 864\,$\mu$m), including the full epitaxial stack (Fig.~\ref{fig:Deflecto}).

\FloatBarrier

This value agrees with in situ curvature measurements indicating a bow of -16.5\,$\mu$m at the end of growth (93$^\circ$C). The initial GaAs/Si template exhibits a concave bow of -8.4\,$\mu$m. Using Stoney’s relation, the resulting VCSEL film stress is estimated at 108\,MPa:

\begin{equation}
\sigma = \frac{4}{3} \times \frac{E}{(1 - \nu)} \times \frac{t_s^2 (B_{f}-B_{i})}{t_f L^2}
\end{equation}

where $\sigma$ is the film stress, $\frac{E}{(1 - \nu)}$ the biaxial modulus, $t_s$ and $t_f$ the substrate and film thicknesses, $B_{f,i}$ the final and initial bow, and $L$ the wafer diameter.

Unlike bulk GaAs on Si, which typically shows convex bowing \cite{Jordan1988}, the observed concave deformation is attributed to thermal effects. After full relaxation at growth temperature, the larger thermal expansion coefficient of GaAs relative to Si leads to concave bowing upon cooling. According to Timoshenko’s model \cite{Timoshenko1925}:

\begin{equation}
\label{eqn:timoshenko}
B = -\frac{3L^{2}}{4} \times \frac{\bar{E}_{f}(\alpha_{f} - \alpha_{s})(T_{growth} - T_{amb})t_{f}}{\bar{E}_{s}t_{s}^{2}}
\end{equation}

A numerical evaluation yields a bow of approximately -30\,$\mu$m for a 2-inch wafer, consistent with experimental measurements.

Deflectometry also provides nanoscale surface information. Filtered data (Fig.~S5) indicate low roughness ($\lesssim 3\,$nm) over coherence lengths exceeding 50\,$\mu$m, with no evidence of crosshatching or hillock formation, in agreement with AFM observations.

\subsection{VCSEL device fabrication and characterizations}

\begin{figure}[htbp]
\centering
\includegraphics[width=1\textwidth]{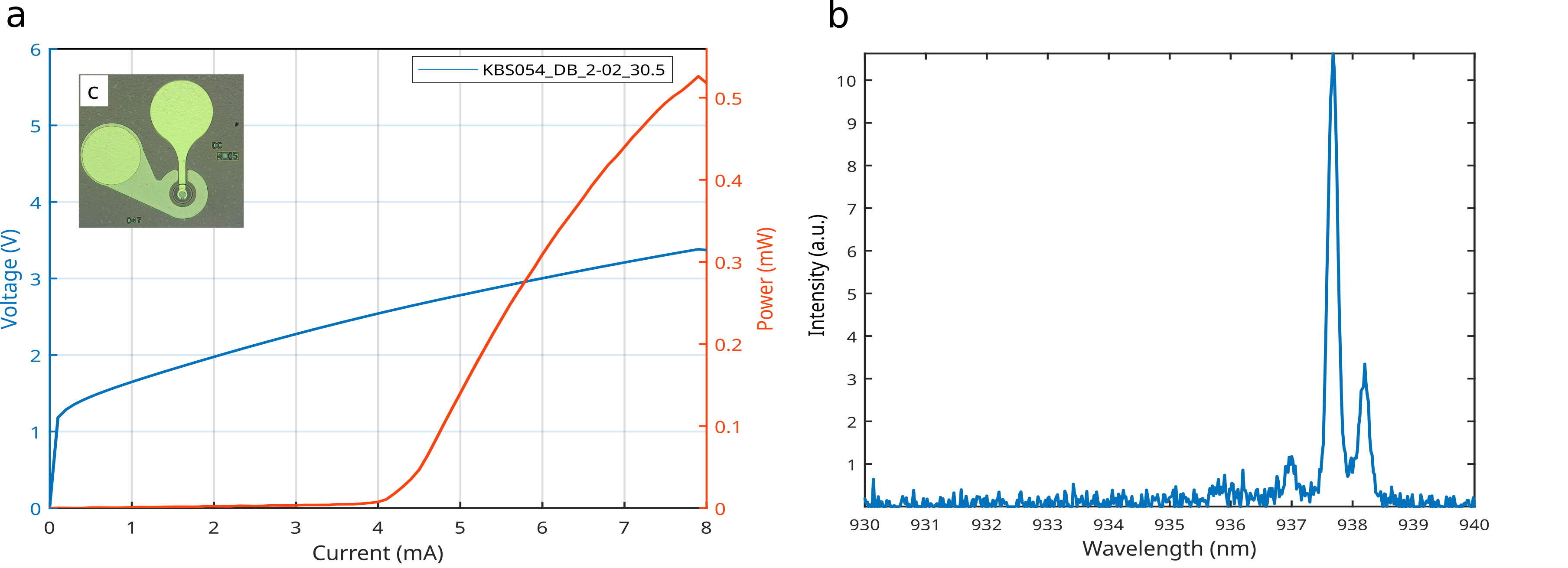}
\caption{(a) Power-current-voltage characteristics at room-temperature under continuous wave operation. (b) Emission spectrum at 5\,mA. (c) Fabricated and measured VCSEL device (oxide aperture $\sim 3-4\,\mu m$.}\label{fig:LIV_spectrum}
\end{figure}

The VCSEL device processing flow is illustrated in the supplementary document (Fig. S2). The VCSEL fabrication steps are typical, involving anode and cathode electrical contacts both on the front epitaxy side, and plasma-etched mesas with various diameters on the processed sample. An oxide aperture is formed by the selective thermal oxidation of a 30\,nm-thick Al$_{0.98}$Ga$_{0.02}$As layer placed close to the first standing-wave field node above the cavity (as seen on fig. \ref{fig:TEM}a,b).    
A final processed device is shown on Fig. \ref{fig:LIV_spectrum}c.

Figures \ref{fig:LIV_spectrum}a and b show the laser performance of the fabricated VCSEL structure with a $3.5\,\mu\mathrm{m}$-wide oxide aperture. A threshold current of 4 mA, and a maximum optical output power of 0.5 mW at thermal rollover are achieved. The electrical characteristics show a kink voltage of the PN diode at 1.18 V, which is typical for AlGaAs-based VCSEL, as well as a differential series resistance of 200\,$\Omega$ at 8\,mA.

\FloatBarrier

The emission spectrum (Fig. \ref{fig:LIV_spectrum}b) shows a multi transverse-mode behavior, which can be explained by the non-circular shape of the oxide aperture.  

While these performances are lower than those of a standard VCSEL grown on GaAs, they largely outdo the previous results reported on electrically-pumped heteroepitaxial AlGaAs-based structures on a Si substrate\cite{Egawa1997, Deppe1990}. Our results also show the benefits of notable improvements of the optical properties of DBRs, compared with recent reports\cite{Guo2023, Laryn2025}. Typically, we measured Si-VCSEL devices exhibiting threshold current densities between 10 and 20\,$kA/cm^2$, corresponding respectively to aperture diameters from 4.5 to 3\,$\mu$m. This can be compared with an equivalent density of 5-10\,$kA/cm^2$ for GaAs-VCSEL with the same aperture range\cite{Choquette1997}. \\

While this initial proof-of-concept demonstration of the direct growth of III-V VCSELs on silicon --- and the demonstration that they can approach the performance of standard VCSELs --- is very promising, significant improvements in the quality of the III-V materials are required to make these Si-based VCSEL components suitable for applications. This would require better control of strain and its relaxation, and uniformity over larger wafer areas. 

VCSELs with larger apertures than 5\,$\mu$m did not demonstrate lasing operation indicating that defect-related non-radiative recombination and optical losses still constrain performance and scalability. 
Considering the TDD of $3\times10^{7}\,\mathrm{cm^{-2}}$ measured by ECCI, we estimate that approximately three dislocations are present within the 3.5\,$\mu$m aperture diameter. These dislocations can impact both the laser performance and the morphology of the oxide aperture. Indeed, we observe a non-uniform oxide aperture shape that deviates markedly from the expected circular or diamond geometry. Threading dislocations can influence the lateral selective thermal oxidation of AlGaAs by providing fast diffusion pathways through surface-emerging defects\cite{Peng2005}, thereby locally modifying the oxidation kinetics and significantly altering the aperture geometry and dimensions.

Also, thermal cracks on the surface of the processed sample decreased the yield of functional lasers. Moreover, resonance was optimally aligned with the MQW gain spectrum only over a portion of the processed sample, so lasing could be achieved in only a limited number of devices.

\section{Discussion}\label{sec13}

In summary, the direct epitaxial growth of high-performance AlGaAs-based VCSELs on silicon substrates, comparatively to the state-of-the-art, has been successfully demonstrated. This work systematically addresses the long-standing challenges associated with lattice mismatch, defect propagation, and strain management in III-V-on-Si integration, with particular emphasis on buffer layer engineering and heterostructure optimization. By combining MOCVD for efficient defect filtering within a thin (<2 $\mu m$) buffer layer and MBE for precise growth of the Bragg mirrors and MQW active region, we establish a growth strategy that preserves both structural integrity and optical efficiency.
Comprehensive in situ and ex situ characterisation confirms effective suppression of threading dislocations, smooth heterointerfaces, and high-quality MQW regions. The implementation of a thick, fully relaxed III-V multilayer VCSEL stack further reduces residual strain and wafer bowing, demonstrating mechanical robustness and scalability toward larger wafer formats. Notably, the resulting devices exhibit enhanced Bragg mirror reflectivity and laser performance metrics not previously achieved for AlGaAs-based VCSELs directly grown on silicon.
Beyond device demonstration, this study establishes a viable and scalable pathway toward monolithic integration of efficient VCSEL sources on silicon. By eliminating the need for native III-V substrates, the proposed approach offers reduced manufacturing complexity, improved compatibility with established silicon photonics platforms, and potential benefits in cost and sustainability.
While further optimization is required to evaluate long-term reliability\cite{Liu2022}, thermal management under continuous-wave operation, and wafer-scale uniformity and reproducibility \cite{Shang2020}, the present results represent a decisive step from material innovation toward manufacturable silicon-integrated VCSEL technology. Collectively, this work advances the foundation for high-density, on-chip photonic systems requiring compact, stable, and high-performance light sources.

\section*{Methods}\label{sec11}

\textbf{Epitaxy}\\
The GaAs/Si growth template was fabricated using InGaAs/GaAs quantum well layers combined with thermal cycle annealing (TCA) on an on-axis Si(001) substrate. The epitaxial growth was carried out in an Applied Materials metal-organic chemical vapor deposition (MOCVD) system.

Prior to epitaxy, a 300\,mm diameter Si(001) substrate, featuring a miscut angle below 0.5$^{\circ}$ toward the <110> crystallographic direction, underwent high-temperature annealing at 900$^{\circ}$C under a hydrogen atmosphere \cite{Martin2016}. To suppress the formation of antiphase boundaries (APBs), we combined the 900$^{\circ}$C annealing with a two-step process in which an 80\,nm GaAs epitaxial layer was grown. The structure was further reinforced by the deposition of a 1\,$\mu m$ GaAs buffer layer without antiphase boundaries at 640$^{\circ}$C.

The template fabrication was completed with the growth of a 20\,nm GaAs capping layer at 640$^{\circ}$C.
Finally, the GaAs-on-silicon template was diced into 2-inch wafers for subsequent molecular beam epitaxy (MBE) growth.

The VCSEL structure was grown on a Riber MBE412 cluster system equipped with two ABN300DF cells for Ga evaporation set at 0.1\,$\mu m/h$ and 0.5\,$\mu m/h$ GaAs growth equivalent growth rates, two ABN150DF cells for Al set at 0.5\,$\mu m/h$ and 0.9\,$\mu m/h$ AlAs growth equivalent growth rates, an ABN150DF cell for In and an ABN151D for Si n-type doping. A $CBr_4$ injector was used for carbon p-type doping. Finally, As was delivered by a VAC500 valved cracker cell in $As_4$ mode (no cracking). After thermal deoxidation at about 630$^{\circ}$C observed by RHEED and under $As_4$ pressure, a 1\,$\mu m$ thick Si-doped GaAs buffer layer was grown at 600$^{\circ}$C at an $As_4$ Beam Equivalent Pressure (BEP) of $6\times10^{-6}\,Torr$ (V/III = 2 at 0.5\,$\mu m/h$). Then As is raised to a BEP = $1.92\times10^{-5}\,Torr$ (V/III = 3.2 at 1\,$\mu m/h$) until the end of VCSEL growth. The substrate temperature is lowered to 520$^{\circ}$C before the growth of the three 7.5\,nm thick $In_{0.1}Ga_{0.9}As$ quantum wells, and then raised  up to 600$^{\circ}$C for the growth of the top Bragg mirror.

\textbf{Material characterizations}\\
In-situ curvature and spectral reflectometry were monitored during MBE growth using in-house systems developed and transferred to Riber, now commercially available as EZ-CURVE and EZ-REF.

Curvature is measured using the magnification-based method \citep{Arnoult2021}. Two optical heads, mounted on CF35 heated viewports and tilted at 70$^{\circ}$ to the wafer normal, face each other. One contains a 3x3 array of white spots from a collimated LED through a perforated plate, while the other supports a camera to image the reflected pattern. Spot spacing is recorded at 100 fps to determine real-time curvature changes, with data synchronized to wafer rotation to provide an average over $\sim$500 points per rotation.

The reflectometry setup uses a 70 W halogen lamp coupled to an optical fiber and an injection head mounted on a CF35 heated viewport at 20$^{\circ}$ from the wafer normal. This head, containing the fiber output and a collimating lens, faces a detection head with an s-polarized linear polarizer. A 50/50 beamsplitter splits the reflected signal into two fibers connected to spectrometers (AvaSpec-ULS2048CL-EVO and NIR256-1.7-EVO), covering 320-1701 nm. Both are synchronized with wafer rotation to minimize signal disturbances.

%\textbf{TEM observations}
TEM images were obtained using a JEOL JEM 2100F transmission electron microscope operated at 200 kV. A CMOS Gatan RIO16IS 4K*4K camera was used for digital acquisition of micro-diffraction images.
The TEM cross-section lamella was prepared using a focused ion beam (FIB)-SEM (Helios 600i, FEI) with the lift-out technique, protected with carbon and platinum layers, and thinned to $\sim$80\,nm.

%\textbf{AFM}
AFM surface characterization was done in tapping-mode on a Bruker ICON instrument to analyze surface topography and roughness.

%\textbf{ECCI}
ECCI was used to quantify the dislocation density on a Gemini 460 Zeiss SEM, wih an insertable backscattered electrons detector (BSD).

%\textbf{Reflectivity measurement by FTIR}
Optical reflectance was measured by Fourier Transform InfraRed spectroscopy (FTIR) using a mapping IR accessory (Pike Technologies). The FTIR model used was a Vertex 70 from Bruker with a tungsten source and Si detector.

%\textbf{Photoluminescence}
Photoluminescence measurements were conducted on an S\&I TriVista bench at 300 K. A frequency-doubled Nd:YAG laser from Cobolt was used as the excitation source at 532 nm. The sample was mounted on motorized stages featuring a 50 nm resolution. The PL emission was collected by a 10x 0.30 NA objective (MPlanFL N, Olympus) and acquired by an Andor InGaAs detector with a 30 l/mm, 1100 nm blazed grating.

%\textbf{X-ray diffraction}
X-ray diffraction analyses were carried out using a Bruker D8 Discover diffractometer equipped with a Cu sealed tube source. The beam is conditioned by a Göbel mirror, a 1 mm wide slit and a two-bounce Ge(004) monochromator (Cu K$\alpha$1), and detected using a one bounce Ge crystal (pathfinder), a 0.1 mm slit and a scintillation counter.

%\textbf{Deflectometry measurement}
Wafer shape measurements were performed using a high-resolution deflectometry system (D50 from DIP-VIEW), providing a field of view of 56 mm, a lateral resolution of 12\,$\mu$m, and sub-nanometer vertical resolution. This instrument is based on a proprietary white-light Phase Modulation Deflectometry (PMD) technique, enabling precise measurement of specular surfaces topography.

\textbf{Device fabrication and characterizations}\\
The processing flow is standard for VCSEL devices fabrication including successively mesa ecthing, thermal oxidation, dielectric passivation and windows opening, and anode and cathode contact deposition.
For mesa etching, a 2.3\,$\mu$m-thick SPR700 photoresist layer was used as hard mask. Cl$_2$/N$_2$-based dry etching by ICP-RIE (Sentech SI500) was applied downto the bottom DBR. Thermal wet oxidation was performed at 430$^{\circ}$C in a furnace equipped with in-situ optical monitoring of the aperture size. 300\,nm SiO$_x$ was subsequently deposited by ICP-PECVD (Oxford, Plasmalab 100), and after a photolithography step, openings were etched by ICP-RIE (Ar/CF$_4$). Metal pads were realized through lift-off process, with Ti/Pt/Au metal stack for P-side and AuGeNi/Au fo N-type contacts.   

%\textbf{Light-current-voltage characteristics and sprectrum measurements}
Light-current-voltage characteristics were measured at room temperature with a probe system (MPI TS2000), under continuous wave current injection (Keysight B1500A semiconductor device analyzer), and light collected using a 27.9\,mm-wide circular Si photodiode (OSIO 25DP). Spectra were acquired with an optical spectrum analyzer (Yokogawa AQ6370D) with 0.1\,nm resolution. 

\backmatter

%\bmhead{Data availability}
%
%The data that support the findings of this study are available from the corresponding author upon reasonable request.

\bmhead{Acknowledgements}

The research was supported by the EU PhotoGeNIC project (GA 101069490), the ''Investissements d'avenir'' program (ANR-15-
IDEX-02) through the LabEx MicroElectronics.
This work was technically supported by the LAAS-CNRS and LTM-CNRS/UGA micro-nanotechnologies platforms within RENATECH network, the CEA-Leti platform, and the joint laboratory EPICENTRE LAAS/RIBER.

%\bmhead{Author contributions}
%The manuscript was written with contributions from all authors. All authors have approved the final version of the manuscript.
%
%\bmhead{Conflict of interest}
%\bmhead{Supplementary information} is available for this paper in a other joint document.

\bibliography{Biblio_Si-VCSEL}% common bib file

\end{document}